\documentclass[letterpaper,twoside,twocolumn,english,prl]{revtex4}
\usepackage[T1]{fontenc}
\usepackage[latin9]{inputenc}
\usepackage{amsmath,xcolor}
\usepackage{esint}
\usepackage{graphicx}
\usepackage{placeins}
\usepackage{float}
\usepackage{subfigure}

\makeatletter

\@ifundefined{textcolor}{}
{%
 \definecolor{BLACK}{gray}{0}
 \definecolor{WHITE}{gray}{1}
 \definecolor{RED}{rgb}{1,0,0}
 \definecolor{GREEN}{rgb}{0,1,0}
 \definecolor{BLUE}{rgb}{0,0,1}
 \definecolor{CYAN}{cmyk}{1,0,0,0}
 \definecolor{MAGENTA}{cmyk}{0,1,0,0}
 \definecolor{YELLOW}{cmyk}{0,0,1,0}
}

\usepackage{babel}

\makeatother

\usepackage{babel}
\begin{document}

\title{Record dynamics in the parking lot model }

\author{Paolo Sibani$^{1}$ and Stefan Boettcher$^{2}$}

\affiliation{$^{1}$FKF, University of Southern Denmark, Campusvej 55, DK5230,
Odense M, Denmark\\
 $^{2}$Department of Physics, Emory University, Atlanta, GA 30322,
USA. }
\begin{abstract}
We present an analytical and numerical study of  the parking lot model (PLM)
of granular relaxation and make a connection to the aging dynamics of dense colloids. 
As we argue, the PLM 
is a Kinetically Constrained Model which features  astronomically large equilibration times
and  displays a characteristic aging behavior on all observable time scales.
The density of parked cars displays quasi-equilibrium Gaussian
fluctuations interspersed by increasingly rare  intermittent events, quakes,  which can lead   to
an   increase of the density to new record values. 

Defining active clusters as the shortest domains
of parked cars which must be re-arranged to allow further insertions,
we find that their typical length  grows logarithmically with time for
low enough temperatures and show how the  number of active clusters  on average gradually 
decreases as the system approaches equilibrium.
We further characterize the aging process
in terms of the statistics of the record sized fluctuations in the interstitial free
volume which lead to quakes and show that quakes are uncorrelated and that
they can be approximately described as a Poisson process in logarithmic time.
\end{abstract}
\maketitle

\section{Introduction\label{sec:Introduction}}

The Parking Lot Model (PLM) is an off-lattice model where identical cars are
 placed on a one-dimensional parking strip with
no marked bays. Its origin  can be traced back to the one dimensional random
packing problem
first considered by Renyi~\cite{Renyi58} decades ago,
where identical objects, i.e. `cars',  are inserted in random positions until no interstitial space
remains which is large enough to accommodate yet another car.
Its (more recent) physical
applications allow both insertion and removal and  include 
 molecule ad- and de-sorption within a crowded surface
area \cite{Krapivsky94b}, and  the  compactification 
of  granular materials \cite{Nowak98}.
Certain  glassy features of the PLM dynamics were  discussed 
in Ref.~\cite{Kolan98} but,  in spite of intense  theoretical focus on
Kinetically Constrained Model   (KCM)
 for their  connections to glassy dynamics~\cite{Ritort03},
 it has  largely  gone unnoticed that the  PLM qualifies as  a KCM.
 
The asymptotic properties of the PLM average observables have 
been explored previously \cite{BenNaim98}. Here we investigate in
more detail its  spatial and temporal complexity, showing 
in particular that   a key property
of glassy dynamics,  
dynamical heterogeneity in time and space,
is  present in this model and  is   related to record sized
fluctuations and their statistics,  as also 
seen  in  other glassy
systems~\cite{Sibani05,Sibani06,Sibani06a,Sibani07,Sibani13}.
Furthermore, the `thermal' model version
presently investigated furnishes  a prime example of  decelerating aging dynamics controlled
by kinematic constraints. Our  analysis clarifies a key model assumption made in     a recent
description of particle motion in dense colloidal suspensions~\cite{Boettcher11,Becker14}.
Specifically, the PLM features reversible fluctuations similar to
in-cage rattlings of dense colloids together with irreversible releases
of free volume.
The latter  are  associated with escapes in a free-energy landscape~\cite{Cipelletti00,Kajiya13}
and are connected to a cooperative and increasingly rare restructuring of the 
spatial domains present in the system.

The basic mechanism behind the model's decelerating dynamics is that
the kinetic constraint provided by car impenetrability becomes  harder
to overcome as the density increases.
A minute and increasingly rare $O(1/N)$-change
to the car density lowers the free energy, but concomitantly raises
the free energy barrier which must be overcome to further increase
the density.
The non-trivial spatial structure  associated to
increasing free-energy barriers \cite{Ritort95}
is indeed responsible for the dynamical behaviour of the PLM: 
let an \emph{active cluster} or \emph{active domain}
be a group as small as possible  of adjacent cars
 which must be
rearranged in order to create an interstitial space sufficient  to accommodate an additional  car.
 As we shall see, the size of active clusters
grows logarithmically with the system's  age, and the characteristic time
for their  rearrangement by means of random moves grows exponentially
with their size, similarly to  what is observed in both recent
experiments~\cite{Yunker09,Zargar13} and in numerical
model simulations~ \cite{Boettcher11,Becker14}  of  aging in 
dense colloids. 

The aging dynamics of the PLM 
 is induced by record fluctuations~\cite{Sibani05,Sibani06,Sibani06a,Sibani07,Sibani13},
 in this case free-energy fluctuations able to bring  the system across  a series of ever increasing  free energy barriers.
 Such fluctuations trigger non-equilibrium \emph{quake}, in our case the rearrangement of an active cluster followed 
by the insertion of an extra car.
Specifically, PLM quakes are  induced by the appearance  of an interstitial volume wide enough to accommodate
one additional car. These fluctuations are rare in dense systems, occur at 
a decelerating rate on a linear time scale but at 
a nearly constant rate 
when viewed on a logarithmic time-scale.
Correspondingly, the model's
dynamics transverses metastable states of growing duration,
each characterized  by reversible fluctuations around a fixed average number
of parked cars and each modified   by an irreversible quake. 

The paper is organized as follows: In the next Section, we introduce
the PLM; in Sec. \ref{sec:Hierarchical-organization}, we analyze
its hierarchical configuration space structure and, using heuristic arguments,
sketch the dynamical consequences this  structure  generically brings;
in Sec. \ref{sec:Thermal-dynamics} we introduce a rejection-less
 Monte Carlo algorithm able to relax the system to equilibrium by transversing
 the required number of metastable states;   in Sec. \ref{sec:Conclusions}
 we 
 make connections with the phenomenology of kinetically
constraint models, draw  some conclusions and offer an outlook on
future work. In the Appendix, we detail
the algorithm used to obtain our numerical results.

\begin{figure*}
\centering
\begin{tabular}{ll}  
   \hspace{-.5cm} \includegraphics[scale=0.4]{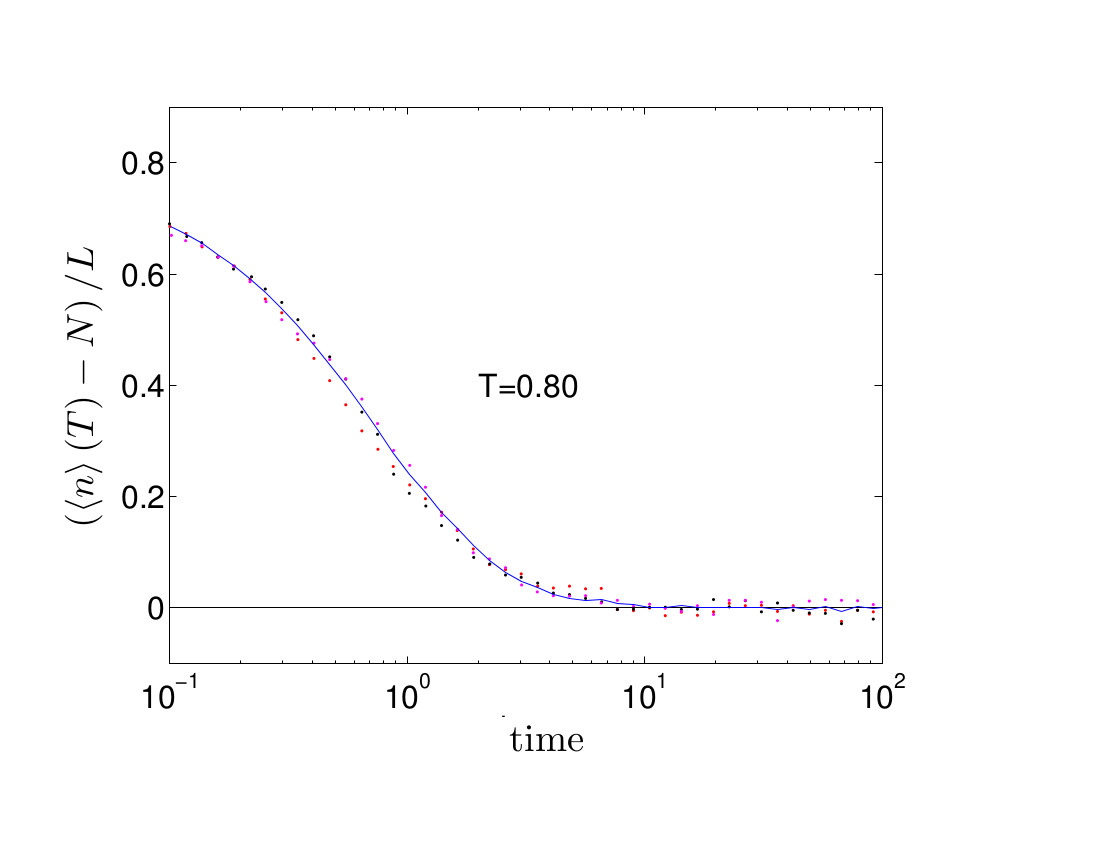} & \hspace{-2.2cm}
      \includegraphics[scale=0.4]{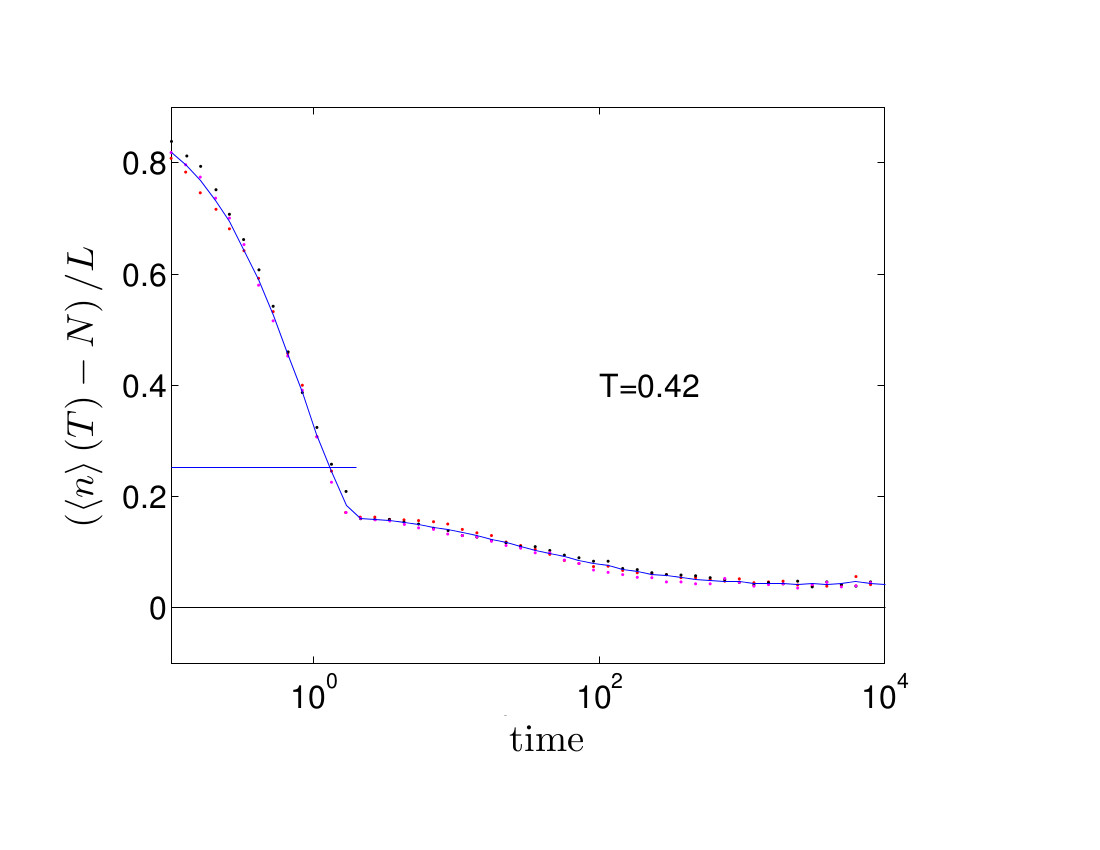}\
  \end{tabular}    
      \caption{In each  panel, the blue line shows  an average over  $20$ independent simulations
      of the deviation of the car density  from its equilibrium  value $(\langle n\rangle(T) -N)/{L}$. All simulations  start from
  an empty lot of length $L=1000$. Three different trajectories are shown as data points to illustrate the fluctuations around the average
  deviation.
      Left-hand panel:  $T=0.80$. The
  ordinate  converges   rapidly  to zero, signaling that  equilibrium  is reached.
  Right-hand panel: $T=0.42$. 
  The horizontal line segment with  ordinate $1-\rho_0$ marks the value of the  at which the 
  $T=0$ greedy  dynamics grinds to a halt.  We can see that our `naive' thermal algorithm comes close 
  to equilibrium, and then remains trapped in a long-lasting metastable state.}
 \label{fig.naive}
\end{figure*}

\section{The Parking Lot Model\label{sec:Description-of-the}}
A brief description of the PLM  is given here   to fix
 the  notation.
Our parking lot is a strip delimited by two rigid walls to avoid center-of-mass
drifts, has no marked bays, and can at most accommodate $L$ cars
of unit width \cite{Renyi58}. At a given time,  $N\leq L$ cars are
present, all parked perpendicularly to the strip's longitudinal axis.
A configuration is equally well specified by a list of $N+1$ interstitial
spaces $I_{i}$, which, for $0<i<N$, separate parked cars $i$ and
$i+1$, with $I_{0}$ separating the left wall from the first car
and $I_{N}$ separating the last car from the right wall. 
We gloss over the distinction between an interstitial
space and its size, or length.

An empty lot has a single interstitial space $I_{0}=L$, and the first
insertion generates two interstitial spaces $I_{0}=q(I_{0}-1)$ and
$I_{1}=(1-q)(I_{0}-1)$, where $q$ is a random number drawn from
the uniform distribution in the unit interval. In general, a new insertion
into an existing interstitial $I_{i}>1$ splits the latter into two
parts. First the indices of the interstitials from $i+1$ and onwards
are incremented by one, and then the $i^{{\rm th}}$ and $i+1^{{\rm st}}$
values are recalculated as $I_{i}\leftarrow q(I_{i}-1)$ and $I_{i+1}\leftarrow(1-q)\left(I_{i}-1\right)$.
For a car removal, we set $I_{i}\leftarrow I_{i}+I_{i+1}+1$, and
decrement by one 
the indices of the interstitials from the rightmost one and down to
$I_{i+2}$.

Starting from an empty lot, random insertions succeed as long as 
interstitial spaces  larger than unity exist.
When this no longer applies, a  `random
loosely packed' configuration is reached which can only be changed 
by two-car processes~\cite{Kolan98}: 
either a `bad parker' is removed leaving sufficient space for two `good parkers', or 
the opposite process occurs.
Such   metastable situation is here  
dubbed~\emph{stage zero} because it turns out to be 
the first in a hierarchy of metastable states.
The average parked car density at  stage zero 
was analytically shown by Renyi \cite{Renyi58} to
approach, for $L\to\infty$,
$\rho_{0}=0.7475\ldots$ ,
a value also close to numerical results obtained for a two dimensional
version of the same problem \cite{Aristoff09}.

In the `thermal' version of the model discussed later in more detail
the basic energy scale is defined by assigning zero energy to parked car and unit energy
to free cars. Hence low temperatures correspond to values
$T\ll1$, 
and the 
`greedy'  random packing algorithm just  mentioned
 corresponds to $T=0$ dynamics,
where only insertion attempts  are possible.
In general, the temperature
$T$ can be so low  and the chemical potential
so high
that any leaving car is  immediately replaced
by another,  the latter  inserted  in the same
slot but with a  slightly shifted  position. A series of such removal/insertion processes thus amounts to
small positional changes of already parked cars, which is similar
to in-cage rattlings of a dense colloid \cite{Weeks00}. 

In each  panel  of Fig.~\ref{fig.naive} the line  represents the average
over $20$ independent simulations of the difference between the equilibrium  
car density  and the  density obtained in simulations 
starting with  an empty lot.
Simulations were conducted using a `naive'  version of the  Waiting Time Method (WTM)~\cite{Dall01}, 
a rejection-less algorithm which inserts and removes cars 
at times   calculated from  the likelihood 
that these moves would  succeed in a standard Metropolis algorithm.
See  also Sec.~\ref{sub:The-Waiting-Time} where a coarser and more efficient version of the WTM
is described.
The left-hand  panel shows 
that at  temperature $T=0.80$
the system equilibrates very quickly.
In the right-hand  panel, the temperature is $T=0.42$, and
the   asymptotic value reached by  the ordinate is   clearly different  from zero,
indicating  the presence of the `zeroth stage' metastable state
mentioned above. 
The  horizontal
line  segment has  ordinate  $1-\rho_0$, and marks the point  at which the $T=0$ greedy dynamics  
on average grinds to a halt. 
As expected,  
the thermal algorithm gets  closer  to  the equilibrium density than the  quench does.
Note the difference in the time scales of the relaxation processes occurring 
at $T=0.80$ and  $T=0.42$. 

\section{Hierarchical Structure
and spatial and temporal heterogeneity\label{sec:Hierarchical-organization}}

In this section the hierarchical structure of PLM configurations
and its relation to spatial and temporal heterogeneity 
is discussed with no mention of  a specific dynamical rule. We do assume, however,  
that the rule in question is `blind', in the sense that a large number
of failed random attempts are needed to successfully rearrange $d$
cars in a preassigned  way. Our conclusions, which are qualitative but general,
 are confirmed in the  next
section, where a numerical approach is   considered.

Defining $I_{{\rm max}}^{0}\ge1$ as the largest interstitial in a
system which has not yet reached metastability, we identify a stage
zero metastable state, $M_{0}(1)$, as the configuration reached when
the condition $I_{{\rm max}}^{0}<1$ becomes fulfilled for the first
time. Such state corresponds to a loosely packed configuration, where
no additional cars can be inserted without previous removals. Using
$k$ independent simulations, each starting from an  empty lot, 
$M_{0}(k),\quad k=1,2,\ldots k_{\rm max}$
loosely packed configurations with the same
statistical properties can be generated,
 which possibly contain
 slightly different numbers of cars. In the limit of large $k_{\rm max}$ and $L$, we
finally obtain 
\begin{equation}
\rho_{0}=\left\langle \frac{M_{0}(k)}{L}\right\rangle _{k}=0.7475\ldots
\end{equation}
for the averaged car density in stage zero, consistent with Renyi's
analytical  result~\cite{Renyi58}.

Even though no additional insertions into any configuration $M_{0}(k)$
are possible, removing the $l^{{\rm th}}$ car will produce enough
space for the insertion of two cars wherever the condition $I_{l}^{1}\stackrel{{\rm def}}{=}I_{l-1}^{0}+I_{l}^{0}>1$
is satisfied. Starting now from a state $M_{0}(k)$ and repeating
whenever possible and as long as possible the random removal of one car followed by the
insertion of two cars in the empty slot thus generated, a stage-one metastable
state $M_{1}(k,1)$ is eventually reached. In such state removing
one car never allows the insertion of two cars, because $I_{{\rm max}}^{1}\stackrel{{\rm def}}{=}\max_{l}\{I_{l}^{1}\}<1$.
Repeating the above 
procedure $m$ times, with   $M_{0}(k)$ as starting point,  and stopping as soon as the condition $I_{{\rm max}}^{1}<1$
is satisfied, generates a series of stage one metastable states $M_{1}(k,m)$.
Each of these can only be modified by randomly searching for two adjacent
cars whose simultaneous removal makes room for three cars. This step  can be iterated until
all possibilities are exhausted.
 Proceeding along this line, we can now define a hierarchy
\begin{equation}
M_{r}(k,m,n\ldots)\subset M_{r-1}(k,m,n,\ldots)\ldots\subset M_{0}(k),
\label{hi}
\end{equation}
where in a  configuration  $M_{r}(k,m,n\ldots)$
 the largest of the sums of all possible sets of $r+1$ adjacent
interstitial spaces  obeys $I_{{\rm max}}^{r}<1$.
The  symbol $\subset$  used in, say,  $M_1(3,0)\subset M_0(0)$
 indicates  that state  $M_1(3,0)$
is  generated dynamically starting from state $M_0(0)$, but does not indicate a static  set inclusion relationship.
The  car density at level $r$ is 
\begin{equation}
\rho_{r}=\left\langle \frac{M_{r}(k,m,n\ldots)}{L}\right\rangle _{k,m,n,\ldots},
\end{equation}
where the average is taken over all the available indices.  
 The critical  car densities separating each of the  first five levels of the hierarchy  from its successor
 were obtained numerically for $L=1000$ and   are given, with $\pm 1\sigma$ error bars, by:
 $\rho_0=0.7476 \pm4\cdot10^{-4}; \rho_1= 0.8587 \pm3\cdot10^{-4}; \rho_2= 0.8992  \pm3\cdot10^{-4};
 \rho_3=0.9205  \pm 2\cdot10^{-4}; {\rm and \;}   \rho_4=0.9343 \pm 2\cdot10^{-4} $. 
  Renyi's result  corresponds  to  the first value  listed.

To obtain a physical process with a proper timescale, we follow Ben-Naim
et al.~\cite{BenNaim98} in assuming that cars move independently,
only constrained by the free volume of interstitial spaces left to
them by their neighbors. Independent car motion translates in turn 
into interstitial spaces $I_l^{r}$, $1\leq l\leq N$, with a marginal
distribution that is uniform but collectively constrained in their total
length, $\sum_{l=1}^{N}I_{l}^{r}=L-N$. In that case, one finds~\cite{Kolan98} that their effective
distribution is asymptotically exponential, 
\begin{equation}
Q(I)\sim\frac{N}{L-N}\, e^{-\frac{NI}{L-N}}=\frac{1}{\langle I \rangle} e^{-\frac{I}{\langle I \rangle}},\qquad L,N\to\infty.\label{eq:expdistr}
\end{equation}
The result is not too surprising, representing merely an exponential
distribution with the mean interstitial length, $\left\langle I\right\rangle =\frac{L-N}{N}$, as its cut-off.
Then, the probability that an interstitial opens up  to
fit in the $N+1^{{\rm st}}$ car of unit size, is given by 
\begin{equation}
p_{N+1|N}(d_r)=\int_{1}^{L-N}dI\, Q(I)\sim e^{-\frac{N}{L-N}}\stackrel{\rm def}{=} e^{-d_r},
\end{equation}
where  
the relation
$\left\langle I\right\rangle d_r = 1$ is used to define 
   the typical size  $d_r$ of the domain in which $r$  cars need to collectively
moved to provide an opening of unit size.

 In a large system, at each level $r$ of the hierarchy many  domains
 $d_r \sim r$
may coexist. Those are the ``soft spots'' where further insertions
are most likely to occur, separating areas that are \emph{minutely}
more resistant to  insertions at this level. 
Dynamical activity will wander from one domain to the next until 
all successful insertions at level $r$ have taken place.
At this point, the domains characteristic
of the next level will start to play  their role.
The spatially localized dynamical activity, which as  we just argued is typical of the PLM, 
  is also the  manifestation
of dynamic spatial  heterogeneity \cite{DH2011}. 

Turning to temporal heterogeneity, or intermittency,
we note that  $p_{N+1}(r_d)$ defines the rate at which the rare
fluctuations occur which trigger a quake, i.e., the demise of a domain of size $d_r$
and the corresponding insertion of an extra car.
Quakes occurring at level $r$ determine the time $\Delta t_{r}$
 it takes to go from the $r^{{\rm th}}$ to the
$r+1^{{\rm st}}$ level of the hierarchy. This time  grows as $t \sim\tau\, e^{d_r}$,
where $\tau$ is a constant.
Conversely, we can say that the size of such domains grows logarithmically
in time, 
\begin{equation}
d_r(t)\sim\ln\frac{t}{\tau}.
\label{log_growth_of_domains}
\end{equation}
The logarithmic growth of the size of such \emph{active domains} is a key property of the cluster model discussed in Ref.~\cite{Becker14} and
also represents a key prediction of the
record dynamics description of colloidal aging~\cite{Boettcher11}.

To further connect  our record dynamics picture with previous work \cite{BenNaim98}
on the PLM, we also note   that 
\begin{equation}
d_r=\frac{1}{\left\langle I\right\rangle }=\frac{N}{L-N}=\frac{\rho_{r}}{1-\rho_{r}},
\end{equation}
where $\rho_{r}=N/L$ is the car density when domains have size $d_r$.
Then, for  times $t$ such that $d_r \gg 1$ and $T\rightarrow 0$ the density approaches $\rho_{\infty}=1$
as 
\[
\rho_{r}(t)=\frac{d_r}{1+d_r}\sim\rho_{\infty}-\frac{1}{\ln\frac{t}{\tau}},
\]
as expected for the PLM \cite{BenNaim98}.

\section{Thermal dynamics\label{sec:Thermal-dynamics}}
To check the dynamical behavior just described in qualitative  terms,
we turn to the  numerical analysis of a `thermal' version of the PLM, where 
parking a car changes its energy
from  $\epsilon=1$ to $\epsilon=0$.
 Cars  are in contact with a thermal energy reservoir at temperature $T$
and, in the lack of interactions, the mean energy per  car and the mean number of parked cars
are given by 
\begin{equation}
\langle \epsilon(T)\rangle=\frac{\exp(-1/T)}{1+\exp(-1/T)}; \quad \langle n(T)\rangle=\frac{L}{1+\exp(-1/T)} \label{eq_N}
\end{equation}
in thermal equilibrium.
In the above, both  temperature and energy are dimensionless
and the Boltzmann constant is set to one.
The kinematic constraint forbidding the spatial
overlap of parked cars has no effect 
on  thermal equilibrium properties
but has a strong effect on the time scale
needed to achieve    thermalization.

To see how the effect comes about, we note that Eq.~(\ref{eq_N}) for $\langle n(T)\rangle=\rho_r L$ defines a series of characteristic temperatures
\begin{equation}
T_{r}=\frac{1}{\ln\left(\frac{\rho_{r}}{1-\rho_{r}}\right)}=\frac{1}{\ln(d_r)},\qquad T_{r}<T_{r-1}\ldots<T_{0},\label{temps}
\end{equation}
each corresponding to the equilibrium  density
at the `edge' between   metastable states $r$ and $r+1$. For $T>T_{r}$, the equilibrium car density 
satisfies $\langle n(T)\rangle/L<\rho_{r}$, and,
consequently, a dynamical process starting from an empty lot reaches
equilibrium  before
reaching a  metastable state of stage
$r+1$ or higher. In particular, for $T>T_{0}\approx 0.921$ the equilibrium car
density is too low for the kinematic constraint to play any
role. 

\begin{figure}
\centering
\begin{tabular}{lcr}   
\hspace{-.6cm}   \includegraphics[width=1.15\linewidth,angle=0]{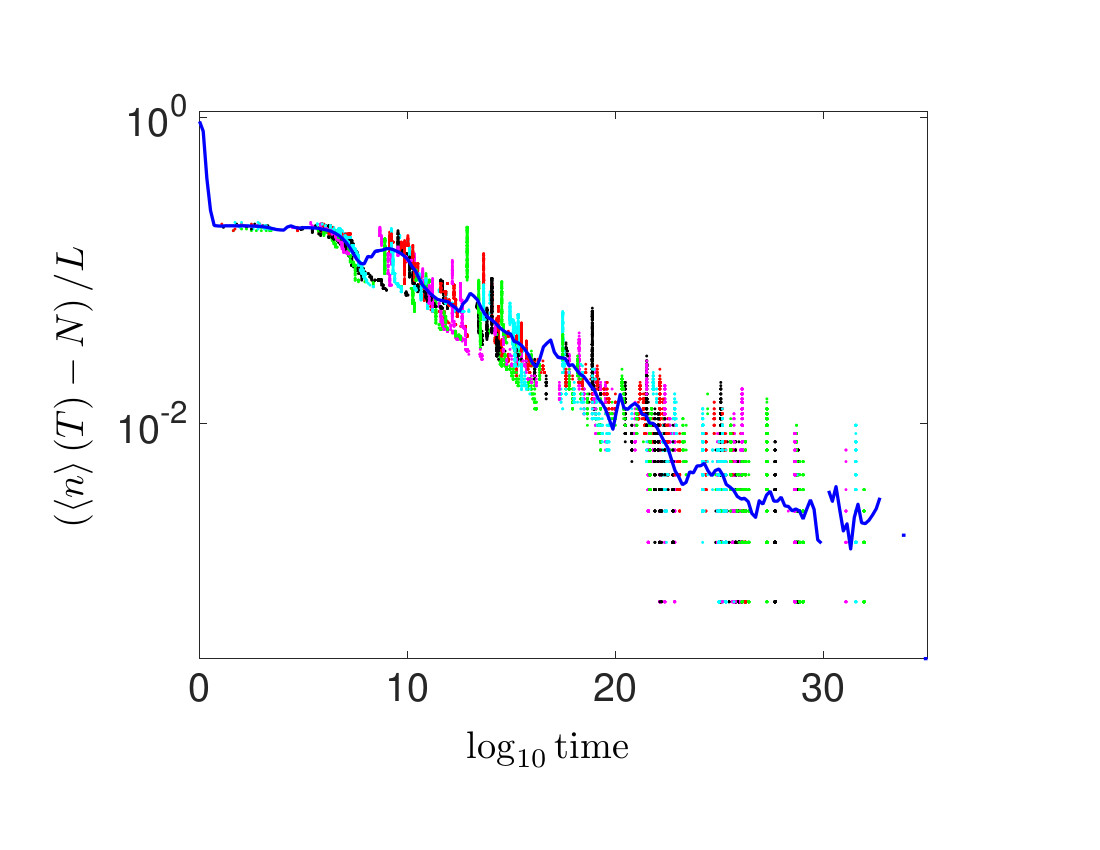}     \\
   \includegraphics[width=\linewidth]{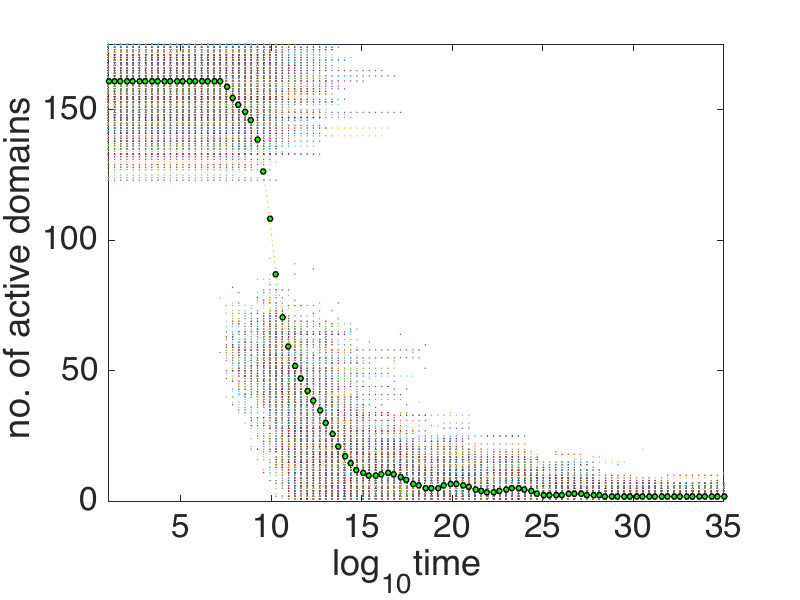}\\
  \end{tabular}    
      \caption{Upper panel: The blue line depicts the average of $2^{10}$ 
      independent trajectories, each consisting of the difference
       between the  equilibrium average of the car density at $T=0.35$, 
      and the time dependent  car density 
       when starting from
  an empty parking lot of length $L=1000$.  The
  ordinate  goes through a fast initial decrease,  followed by a considerably slower
 relaxation toward zero. To convey  an idea of the size of the fluctuations,
  $40$ trajectories are depicted as  dots. Negative data values are present in the
  late stages of the relaxations and are omitted.
  Lower panel: the dots show the  number of active domains present is the system versus the logarithm of time 
   for $2^{10}$ independent simulations. 
  The green  circles depict averages over the data points..
  }
 \label{fig.smart}
\end{figure}

 The  equilibrium thermal  density at $T=0.8$ , $\langle \rho \rangle= .777$, is  only slightly above the 
Renyi density and, as shown in the  left panel of Fig~ \ref{fig.naive}, the calculated  discrepancy   $\langle \rho \rangle - N/L$ 
quickly  equilibrates and vanishes.
The right panel shows corresponding data for $T=0.42$, where 
 the 
equilibrium thermal  density, $\langle \rho \rangle= .966$  is way above the
Renyi density and where no equilibrium is reached 
within the time scales considered.

In order to equilibrate starting from an empty lot, a system quenched to a  
low temperature must surmount a number of growing free energy barriers.
This forms the basis of the PLM aging dynamics since, as shown below, the
equilibration time $t_{{\rm eq},r}$ can easily outlast the patience of any observer.
To estimate $t_{{\rm eq},r}$ at temperature $T_{r}$, we use Eqs.~\eqref{log_growth_of_domains}
and \eqref{temps}, finding, for an unspecified  numerical constant $C$, 
\begin{equation}
t_{{\rm eq},r}=\exp(C\exp(1/T_{r})),\label{eq_time}
\end{equation}
a quantity which becomes astronomically large when $\rho_{r}\rightarrow1$
and $T_{r}\rightarrow0$.

To explore the thermal aging dynamics of the PLM, the
Metropolis algorithm is woefully inadequate, since it would spend
most computer resources to generate and reject configurations.
We use instead an adaptation of the Waiting Time Method
(WTM) \cite{Dall01}, a rejection-less algorithm whose application
to the PLM is sketched in the Appendix. 
The key points are: \emph{i)} the algorithm generates for each system state  a list of possible 
moves, each associated with a time at which the move would happen in 
a sequence of random attempts. The move with the shortest waiting time is  carried out. \emph{ii)}
The algorithm use the temperature value   needed to equilibrate the system at that temperature, see e.g.
 the upper panel of Fig.~\ref{fig.smart}. 
 
At the zero'th stage,  car insertions do not require  prior removals,
and the algorithm inserts and removes single cars with the
frequencies required to approach equilibrium.  
At low temperatures equilibrium densities are high, and the equilibration process
will reach  a stage  where  at least $r$  contiguous cars need to be rearranged  to 
increase the density. Such minimal cluster of cars correspond to the active domains 
$d_r$ defined in Sec.~\ref{sec:Hierarchical-organization}.
  When $r>0$, 
   removing a single car is in most cases followed by a re-insertion
   in the same slot at a slightly different position.
   As argued, this amounts to
quasi-equilibrium fluctuations within the metastable state. 
These fluctuations  are bypassed in our numerical algorithm 
   using two steps:
first, the time needed to 
remove   $r$ contiguous car using  blind attempts
is drawn from an exponential  distribution whose average is proportional to 
the Arrhenius time $\exp(r/T)$.
An  active domain is randomly chosen among those available,
and its  cars   are all removed. This creates a void and 
pushes the system back into its  zero' th dynamical stage.
In the second step, 
 zero' th stage dynamics is utilized, until 
 stage $r'$ with $1 \le r' \le r+1$ is reached.  If
 $r'=r+1$ an extra car has been inserted the event is registered as 
 a  quake, while 
 if  $r'<r$ the system simply enters a lower active stage. In both cases,
  the steps just described are repeated \emph{ad libitum}. 
 Note that, once the system is near  thermal equilibrium,
 the insertion of additional cars through the  zero'th stage dynamical
 step becomes  unlikely and the dynamics enters a fluctuation 
 regime where active clusters of size near the equilibrium cluster size 
 are continuously removed and recreated.
 
The data shown in the  upper panel of Fig.~\ref{fig.smart} are based on 
the differences between the equilibrium car density $\langle n(T)\rangle/L=0.9656$ at $T=0.35$
and the calculated car density $N/L$ at the same temperature 
 for
$2^{10}$  independent simulations,  all starting with an empty lot.
The continuous blue line shows the average value of the differences, and the dots
show  appr. $40$ of our data sets to give an idea of the fluctuations while keeping the 
figure uncluttered. The negative fluctuations  present in the final stages of the 
relaxation are omitted in order to be able to use a vertical logarithmic scale.
The initial phase of the relaxation ends when the  density reaches the Renyi value
$ \rho_0=0.7475\ldots $, i.e., the value  which delimits the lower boundary of the first metastable state.
What then  follows is, on average,  a slow decay of the ordinate toward its equilibrium value, i.e.,  zero.
The equilibration process can also  be followed by 
monitoring the number of active domains.
In the lower  panel of the same figure, the number of   active  domains present at a given time 
is extracted from the same set of simulations
and plotted as dots versus the logarithm of time. The  circles represent the average number
of domains at a given time.

\begin{figure*}
\centering
\vskip 0cm
\begin{tabular}{lr}  
    \includegraphics[scale=0.4]{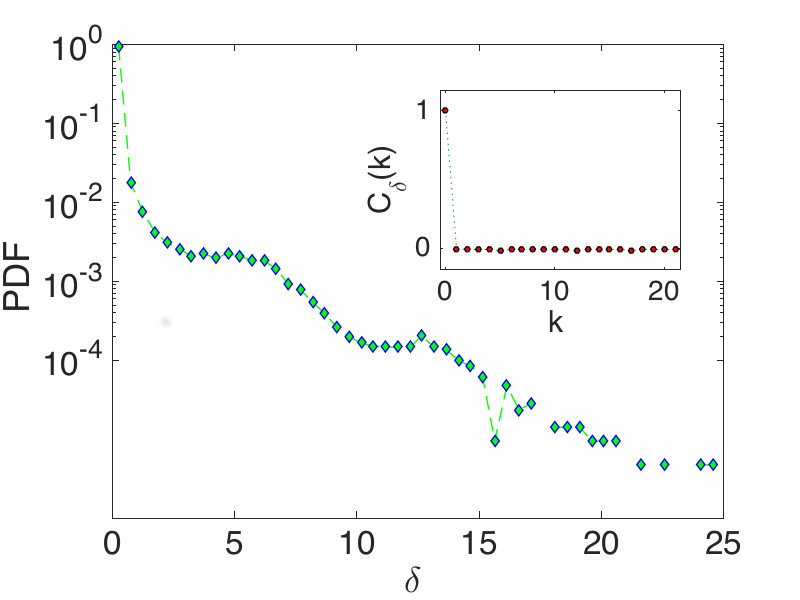} & 
     \includegraphics[scale=0.4]{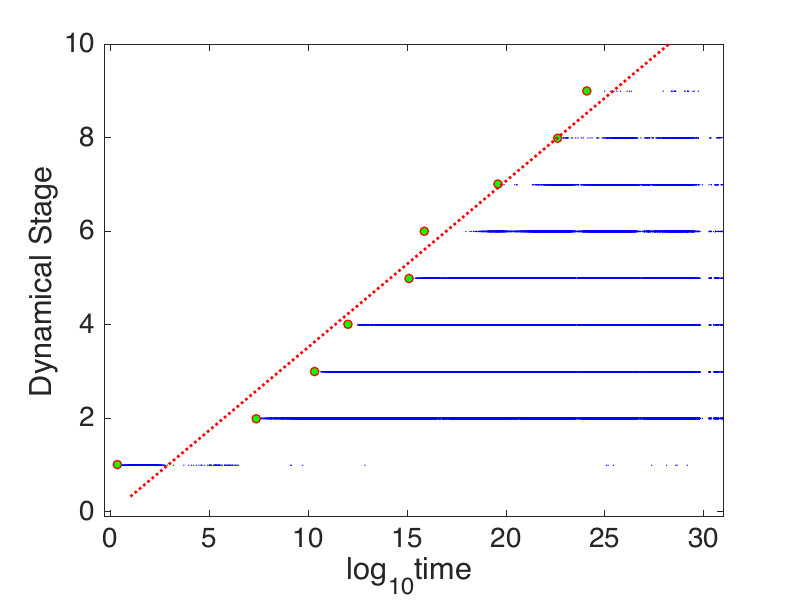}
  \end{tabular}    
      \caption{Left: The main figure shows the PDF  of the differences
      $\delta_k = \log t_k -\log t_{k-1}$, where $t_k$ marks the occurrence of the $k$'th quake.
      The data are extracted from 
        $1024$  independent trajectories run at $T=0.35$, all starting from
  an empty parking lot of length $L=1000$.   The insert shows that the correlation 
  function of the series $\delta_1, \delta_2 \ldots$ is a Kroneker delta indicating
  that successive quakes are independent events.
   Right: The length of the active domains present in the systems
   defines the dynamical stage of the system and  is plotted vs.
  the logarithm of time using blue points. In any small time interval,
  domains of different sizes are generated in rapid succession, giving the false impression
  that domains of different length can coexist.
  The  circles mark the shortest time at which an active domain of a certain length first appear,
  and the red line is linear fit of the data vs. the logarithm of time.
  }
 \label{fig.rd}
\end{figure*}

Figure~\ref{fig.rd}  illustrates how record dynamics predictions fit the low $T$ dynamics of the PLM,
based on estimates  obtained from  our $2^{10}$ independent runs.
Let  $t_k$ denote the time at which the $k$'th quake  occurs, 
and define the `logarithmic waiting times' as the differences
$\delta_k=\log t_k -\log t_{k-1}$.
In a  Poisson process with average $\mu_q\propto \ln t$, 
the logarithmic waiting times are
 independent and exponentially distributed stochastic variables.
 The insert in the 
left panel of the figure shows that different $\delta_k$   have correlation $C_\delta(k) = \delta_{k0}$, indicating
 the required statistical independence. The main figure shows that  the  PDF of the log-waiting times has 
an exponential trend with a superimposed structure not  imputable to statistical
fluctuations.
The average number of quakes (not shown) grows logarithmically in time with a small superimposed
oscillation. In summary,  the quake process is structurally somewhat richer than  a log-Poisson process,
but the latter provides a reasonable simplified statistical description of the salient events of the dynamics.

The length of the active domains marks the dynamical stage reached by  the system and
 is plotted in the right panel  of the figure vs. the logarithm of time.
Active domains of many different sizes replace each other  in rapid succession and
their seeming   co-existence at the same time  
is only due to insufficient graphical resolution.
Longer and longer active domains are seen to develop as the system ages
and the first time they appear is marked by 
the circle at the leftmost edge of every plateau.
The same circle also marks
an   increase in the level of metastability 
or  dynamical stage of  the system.
The red dotted line is a fit of the position of such  events vs. the logarithm of time.
The  clear  logarithmic trend is in excellent agreement with  Eq.~\eqref{log_growth_of_domains}.

\section{Conclusions and outlook\label{sec:Conclusions}}
Kinetically constrained models  have simple equilibrium statistical 
mechanical  and thermodynamical property.
However, an equilibrium or steady-state 
state  can be   hard to reach
since many dynamical paths
in their configuration spaces are blocked 
by kinematic constraints.
The PLM
is a \emph{bona fide} kinetically constrained model,
a fact not  prominently featured in   its origin and history.
The constraint,  no overlaps allowed in
the parking lot, becomes increasingly hard to overcome 
as the density of parked cars increases. This
leads to a rich  aging dynamics, which coexists with  a completely trivial thermodynamics.
The PLM's  equilibration time   grows super-exponentially 
as a function of the 
inverse temperature,  a non-Arrhenius relaxation behavior which
matches the  cooperative nature of the  moves required to relax the system.
 Equilibration is only achievable using 
a rejection-less algorithm which can access to the required time scales
by coarse-graining away 
all  quasi-equilibrium fluctuations. What remains is 
a series of 
 heterogeneous and intermittent non-equilibrium events, connected to the rearrangements of
 active  domains of contiguous cars needed to  insert of an additional car.
These events require climbing free energy barriers of growing height and 
can be approximately described as a
Poisson process with average proportional to $\ln t$, a property which, as discussed in Ref.~\cite{Sibani03} 
 is tantamount to pure aging behavior.

Let us connect our  present findings to  dense colloidal suspensions as described
in Refs.~\cite{Boettcher11,Becker14}, where  key experimental properties 
are reproduced  by a `cluster model' based on the idea that  colloidal particles
belong to clusters whose collapse controls  all irreversible movements in the systems.
To describe an aging colloid, the probability density $P(h)$ 
 that a cluster of size $h$ collapse must decrease very quickly, e.g. exponentially,  with $h$.
 The origin of the clusters and of the form of $P(h)$
  is however left unexplained in Refs.~\cite{Boettcher11,Becker14}.
   Irreversible particle motion was analyzed experimentally by Yunker et al.~\cite{Yunker09} 
  who  defined irreversible events as those   which disrupt  at least three nearest neighbour relationships.
   These authors  find that,  as the system ages,  irreversible changes require the correlated motion
   of  increasingly large clusters. The similarity with the active domains of the PLM
   is clear, since in order to introduce an extra car
   we need the cooperative motion of an increasingly large domain. The probability that a PLM domain be re-arranged
   hence  decreases exponentially with its size, a property which is  shared by the cluster model and 
   which was already  used by 
   Adam and Gibbs~\cite{Adam65} to describe the approach to the glass transition.
   Identifying a PLM  domain with  a cluster of correlated particles in a colloid  points to a statistical mechanism
   shaping  the form of $P(h)$. Stability increases with cluster size as irreversible events are connected
   to a correspondingly decreasing free volume or, equivalently, to a local increase in density. The repulsive short range interactions between 
   colloidal particles prevent such irreversible events  from happening unless a spontaneous collective  fluctuation of $h$ particles 
   provides the free space needed. Such fluctuation becomes exponentially unlikely.

Our analysis indicates that the typical length scale of the domains which have to be rearranged 
in order to approach equilibrium  grows logarithmically with time.
Once equilibrium is reached,  the  typical size of domains
will be larger the longer the equilibration process.
Hence the average domain size will rapidly increase with decreasing temperature. The intimate relation between the two properties is clear in the context of the PLM, 
but could possibly be more generally valid when approaching the glass transition. The issue can  be
investigated experimentally by studying the persistency of  neighborhood  relations in  particle clusters
and its resolution would shed light on the nature of the glass transition.

  Finally, it seems reasonable to speculate  that a similar analysis would apply to other familiar models of slow relaxation and jamming, such as the East model~\cite{Sollich03} or the Backgammon model~\cite{Ritort95}. In the East model, an entire domain of unfrustrated spins has to collectively activate to dislodge and move a single frustrated spin on its boundary, merely to be able to expand by a minute increment. Similarly, in the Backgammon model, $N$ uncoupled particles are spread over $n$ domains, where the energy of the system is proportional to $n$. Particles hop randomly between domains until, by some chance fluctuation of size $\sim N/n$, a domain empties out and becomes inaccessible, leaving $n-1$ domains, each of minutely larger occupation on average.  Thus, these models  share the same phenomenology of clusters of variables requiring  ever new records in the size of collective activations that are exponentially unlikely in their size to progress towards a marginally improved state.

\paragraph*{Acknowledgements:}
SB thanks SDU for their hospitality and the V. Kann Rasmussen Foundation
for supporting the visit. SB further acknowledges financial support from the U. S. National Science Foundation through grant DMR-1207431.
PS enjoyed  interesting discussions with Mads Elmelund Hjadstrup Hansen and Jakob Rath Mortensen.
We'd like to thank an anonymous referee for an interesting suggestion.

\section*{Appendix: }
\subsection{The Waiting Time Method\label{sub:The-Waiting-Time}}

The gist of the WTM is to determine the possible moves in a given
situation, draw for each of these a waiting time from an exponential
distribution with a suitable average, and carry out the move with
the shortest waiting time. The WTM satisfies detailed balance and
eventually generates the Boltzmann equilibrium distribution but, at
low temperatures, does so much faster than the Metropolis algorithm.
The algorithm is particularly simple to apply to the PLM, whose degrees
of freedom have no mutual interactions.

Our version of the algorithm generates a stochastic series $t_{0}<t_{1}<t_{2}<\ldots$
recording the times at which the system configuration undergoes a
change. Depending on the  severity of the constraints, the algorithm
goes through several incarnations, or `stages'. At stage zero, interstitials
are available which can accommodate a car, while in the $k$'th stage,
$k=1,2,\ldots L-2$, a domain consisting of $k$ contiguous cars must
be rearranged, in order to create the space for an additional car.
To reach thermal equilibrium for $T>T_{0}$, only the zeroth stage
of the algorithm is needed while for temperatures in the range $T_{k-1}>T>T_{k},\quad k>1$,
$k$ stages are required.

Initially, $t_{0}=0$, the lot is empty, and the zeroth stage of the
algorithm is applied: Each car is assigned a waiting time $\tau_{i}^{{\rm free}},\quad i=0,1\ldots(L-1)$,
drawn from the exponential distribution with unit average. The car
with the lowest waiting time, say $\tau_{0}^{{\rm free}}$, is selected
for a change of status to `parked', the global time is updated to
$t_{1}=t_{0}+\tau_{0}^{{\rm free}}$ and the waiting times of the
cars which remain parked are synchronized to $t_{1}$, i.e., $\tau_{i}^{{\rm free}}\leftarrow(\tau_{i}^{{\rm free}}-\tau_{0}^{{\rm free}}),\quad i=1,2\ldots(L-1)$.
To complete the first update, the newly parked car is assigned a waiting
time $\tau_{0}^{{\rm parked}}$, drawn from the exponential distribution
with average $e^{1/T}$.

Subsequent updates follow the same pattern as above: time $t_{n}$
is obtained from $t_{n-1}$ by adding the shortest available waiting
time; all other waiting times are synchronized to $t_{n}$, and a
new waiting time for the last car moved is drawn from an exponential
distribution whose average is either $1$ or $e^{1/T}$. The first
choice applies if the last move was a car removal, and the second
if it was a car placement.

As mentioned, for $T_{1}<T<T_{0}$, the dynamics thermalizes in a
metastable state of type $M_{0}$, where insertions are by definition
impossible without previous removals.

With the previous scheme, a car removal would with high probability
be followed by a re-insertion in the same slot, since this is the
only possible sequence unless the sum of the two interstitials adjacent
to the car removed is larger than one. Removal/re-insertion sequences
constitute the bulk of the pseudo-equilibrium fluctuations in the
metastable state but do not change the number $N$ of parked cars,
and do not further the equilibration process. Rather than waiting
for a car removal which allows the placement of two cars to happen
by chance, stage one makes the move and draws the waiting time associated
to it from an exponential distribution, whose average is calculated
as follows: Let $n_{0}$ denote the number of pairs of adjacent interstitials
with total length larger than one (note that $n_{0}>0$ in a metastable
state of type $M_{0}$) and define the above average as 
\begin{equation}
\mu_{0}(n_{0})=\frac{N}{n_{0}}e^{1/T}.
\end{equation}
The first term on the right hand side of the equation is the
average number of random removals needed to select a car surrounded
by one out of  $n_{0}$ interstitial pairs. The second
is the Arrhenius factor associated with its removal. Once the move
is carried out and the global time is updated, the algorithm returns
to the stage zero  update, which continues until a new metastable
state of type $M_{0}$ is identified.

Stage $k$ dynamics entails reshuffling a domain of $k$ adjacent
cars. This is done by first removing the respective cars, and by then
returning to stage zero to fill up the opening thus created. The waiting
time for removing $k$ adjacent cars is drawn from an exponential
distribution, whose average is taken to be 
\begin{equation}
\mu_{0}(n_{k})=\frac{1}{n_{k}}{{N} \choose {k}}e^{k/T},
\end{equation}
where $n_{k}$ is the number of domains of length $r_{k}$ present
in the system. The initial factor accounts for the number of choices
for the placement of the domain, the binomial coefficient is the average
number of attempts needed to place $k$ cars out of $N$ in contiguous
positions, and the exponential is the Arrhenius factor corresponding
to the removal of a group of $k$ cars. As was the case for $k=1$,
the algorithm returns to stage zero to fill in the empty space left
by the removal. We note that removing $k$ cars does not guarantee,
for $k>1$, that $k$ cars can be successfully re-inserted in the
vacant space.
\bibliographystyle{unsrt}
\bibliography{Boettcher,Paolo}

\begin{thebibliography}{10}

\bibitem{Renyi58}
A.~R\'{e}nyi.
\newblock {On a One-Dimensional Problem Concerning Random Space-Filling.}
\newblock {\em {Publ. Math. Inst. Hung. Acad. Sci. }}, 3:109--127, 1958.

\bibitem{Krapivsky94b}
P.~L. Krapivsky and E.~Ben-Naim.
\newblock Collective properties of adsorption-desorption processes.
\newblock {\em J. Chem. Phys.}, 100:6778, 1994.

\bibitem{Nowak98}
Edmund~R. Nowak, James~B. Knight, Eli Ben-Naim, Heinrich~M. Jaeger, and
  Sidney~R. Nagel.
\newblock Density fluctuations in vibrated granular materials.
\newblock {\em Phys. Rev. E}, 57:1971--1982, 1998.

\bibitem{Kolan98}
Amy~J. Kolan, Edmund~R. Nowak, and Alexei~V. Tkachenko.
\newblock Glassy behaviour of the parking lot model.
\newblock {\em Phys. Rev. E}, 59:3094--3099, 1998.

\bibitem{Ritort03}
F.~Ritort and P.~Sollich.
\newblock Glassy dynamics of kinetically constrained models.
\newblock {\em Advances in Physics}, 52:219--342, 2003.

\bibitem{BenNaim98}
E.~Ben-Naim, J.B. Knight, E.R. Nowak, H.M. Jaeger, and S.R. Nagel.
\newblock Slow relaxation in granular compaction.
\newblock {\em Physica D: Nonlinear Phenomena}, 123:380 -- 385, 1998.

\bibitem{Boettcher11}
Stefan Boettcher and Paolo Sibani.
\newblock Ageing in dense colloids as diffusion in the logarithm of time.
\newblock {\em Journal of Physics: Condensed Matter}, 23(6):065103, 2011.

\bibitem{Becker14}
Nikolaj Becker, Paolo Sibani, Stefan Boettcher, and Skanda Vivek.
\newblock Temporal and spatial heterogeneity in aging colloids: a mesoscopic
  model.
\newblock {\em J. Phys.: Condens. Matter}, 26:505102, 2014.

\bibitem{Cipelletti00}
Luca Cipelletti, S.~Manley, R.~C. Ball, and D.~A. Weitz.
\newblock Universal aging features in the restructuring of fractal colloidal
  gels.
\newblock {\em Phys. Rev. Lett.}, 84:2275--2278, 2000.

\bibitem{Kajiya13}
Tadashi Kajiya, Tetsuharu Narita, Veronique Schmitt, Francois Lequeuxa, and
  Laurence Talini.
\newblock Slow dynamics and intermittent quakes in soft glassy systems.
\newblock {\em Soft Matter}, 9:11129, 2013.

\bibitem{Ritort95}
Felix Ritort.
\newblock Glassiness in a model without energy barriers.
\newblock {\em Physical Review Letters}, 75:1190, 1995.

\bibitem{Yunker09}
P.~Yunker, Z.~Zhang, K.~B. Aptowicz, and A.~G. Yodh.
\newblock Irreversible rearrangements, correlated domains, and local structure
  in aging glasses.
\newblock {\em Phys. Rev. Lett.}, 103:115701, 2009.

\bibitem{Zargar13}
Rojman Zargar, Bernard Nienhuis, Peter Schall, and Daniel Bonn.
\newblock Direct measurement of the free energy of aging hard sphere colloidal
  glasses.
\newblock {\em Phys. Rev. Lett.}, 110:258301, 2013.

\bibitem{Sibani05}
P.~Sibani and H.~Jeldtoft Jensen.
\newblock Intermittency, aging and extremal fluctuations.
\newblock {\em Europhys. Lett.}, 69:563--569, 2005.

\bibitem{Sibani06}
Paolo Sibani.
\newblock Mesoscopic fluctuations and intermittency in aging dynamics.
\newblock {\em Europhys. Lett.}, 73:69--75, 2006.

\bibitem{Sibani06a}
{P. Sibani, G.F. Rodriguez and G.G. Kenning}.
\newblock Intermittent quakes and record dynamics in the thermoremanent
  magnetization of a spin-glass.
\newblock {\em Phys. Rev. B}, 74:224407, 2006.

\bibitem{Sibani07}
{P. Sibani}.
\newblock {Linear response in aging glassy systems, intermittency and the
  Poisson statistics of record fluctuations}.
\newblock {\em Eur. Phys. J. B}, 58:483--491, 2007.

\bibitem{Sibani13}
Paolo Sibani.
\newblock Coarse-graining complex dynamics: Continuous time random walks vs.
  record dynamics.
\newblock {\em EPL}, 101:3004, 2013.

\bibitem{Aristoff09}
David Aristoff and Charlesi Radin.
\newblock { Random Loose Packing in Granular Matter}.
\newblock {\em {Journal of Statistical Physics}}, 135:1--23, 2006.

\bibitem{Weeks00}
Eric~R. Weeks, J.C. Crocker, Andrew~C. Levitt, Andrew Schofield, and D.A.
  Weitz.
\newblock Three-dimensional direct imaging of structural relaxation near the
  colloidal glass transition.
\newblock {\em Science}, 287:627--631, 2000.

\bibitem{Dall01}
Jesper Dall and Paolo Sibani.
\newblock {Faster} {M}onte {C}arlo simulations at low temperatures. {T}he
  waiting time method.
\newblock {\em Comp. Phys. Comm.}, 141:260--267, 2001.

\bibitem{DH2011}
Ludovic Berthier, Giulio Biroli, Jean-Philippe Bouchaud, Luca Cipelletti, and
  Wim van Saarloos, editors.
\newblock {\em Dynamical Heterogeneities in Glasses, Colloids, and Granular
  Media}.
\newblock Oxford University Press, 2011.

\bibitem{Sibani03}
Paolo Sibani and Jesper Dall.
\newblock {Log-Poisson statistics and pure aging in glassy systems.}
\newblock {\em Europhys. Lett.}, 64:8--14, 2003.

\bibitem{Adam65}
Gerold Adam and Julian~H. Gibbs.
\newblock On the temperature dependence of cooperative relaxation properties in
  glass-forming liquids.
\newblock {\em J. Chem. Phys.}, 43:139, 1965.

\bibitem{Sollich03}
P.~Sollich and M.~R. Evans.
\newblock Glassy dynamics in the asymmetrically constrained kinetic {{Ising}}
  chain.
\newblock {\em Phys. Rev. E}, 68:031504, 2003.

\end{thebibliography}

\end{document}